\documentclass[a4paper,11pt]{article}

\usepackage[nointlimits]{amsmath}
\usepackage{amsfonts}
\usepackage{amssymb}
\usepackage{amsfonts}
\usepackage{amssymb}
\usepackage{amsthm}
\usepackage{amsmath}
\usepackage{latexsym}
\usepackage{graphicx}
\usepackage{layout}
\usepackage[english]{babel}
\numberwithin{equation}{section}

\usepackage{graphics}
\usepackage{epsfig}

\newtheorem{lemma1}     {Lemma}[section]
\newtheorem{teorema1}   [lemma1]{Theorem}
\newtheorem{prop1}      [lemma1]{Proposition}
\newtheorem{coroll1}    [lemma1]{Corollary}
\newtheorem{cong1}      [lemma1]{Conjecture}
\newtheorem{remark1}    [lemma1]{Remark}
\newtheorem{defin1}     [lemma1]{Definition} 
\newtheorem{example1}   [lemma1]{Example} 

\newenvironment{Lemma}[1][]
        {\begin{lemma1}[#1]\begin{samepage}}{\end{samepage}\end{lemma1}}
\newenvironment{Theorem}[1][]
        {\begin{teorema1}[#1]\begin{samepage}}{\end{samepage}\end{teorema1}}
\newenvironment{Proposition}[1][]
        {\begin{prop1}[#1]\begin{samepage}}{\end{samepage}\end{prop1}}
\newenvironment{Corollary}[1][]
        {\begin{coroll1}[#1]\begin{samepage}}{\end{samepage}\end{coroll1}}

\newenvironment{Remark}[1][]
        {\begin{remark1}[#1]\begin{samepage}}{\end{samepage}\end{remark1}}
\newenvironment{Definition}[1][]
        {\begin{defin1}[#1]\begin{samepage}}{\end{samepage}\end{defin1}}
\newenvironment{Example}[1][]
        {\begin{example1}[#1]\begin{samepage}}{\end{samepage}\end{example1}}

\newcommand{\nada}[1]   {}

\newcommand{\area}         {\mathcal S}

\newcommand{\C}         {\mathcal C}

\newcommand{\energy}       {E}
\newcommand{\eps}       {\varepsilon}

\newcommand{\grad}      {\nabla}

\newcommand{\lagdens}   {\mathcal L}

\newcommand{\N}         {\ensuremath{\mathbb N}}
\newcommand{\normvel}         {v}

\newcommand{\paramgiando}        {s}
\newcommand{\paramgiandoa}        {y}
\newcommand{\paramgiandox}        {x}
\newcommand{\paramspace}        {x}
\newcommand{\paramtemp}        {t}

\newcommand{\rightextremum}       {L}
\newcommand{\R}         {\ensuremath{\mathbb R}}
\newcommand{\Rn}        {\ensuremath{\mathbb R^n}}
 
\newcommand{\spaceimm}       {\gamma} 
\newcommand{\spacetimeimm}       {X} 
\newcommand{\tangvect}       {\frac{\gamma_x}{\vert \gamma_x\vert}} 
\newcommand{\varmuta}       {\sigma} 
\newcommand{\weaksol}       {\spaceimm}

\begin{document}

\title{Closure and convexity results for closed relativistic strings}

\author{
Giovanni Bellettini\footnote{
Dipartimento di Matematica,
Universit\`a di Roma Tor Vergata,
via della Ricerca Scientifica, Roma 00133, Italy,
and
Laboratori Nazionali di Frascati dell'INFN,
via E. Fermi, 40, Frascati (Roma) 00044, Italy.
e-mail: Giovanni.Bellettini@lnf.infn.it}
\and
Jens Hoppe\footnote{Department of Mathematics,
Royal Institute of Technology,  10044 Stockholm, Sweden.
e-mail: hoppe@kth.se}
\and
Matteo Novaga\footnote{
Dipartimento di Matematica Pura e Applicata,
Universit\`a di Padova,
via Trieste 63, Padova 35121, Italy.
e-mail: novaga@math.unipd.it}
\and
Giandomenico Orlandi\footnote{
Dipartimento di Informatica, Universit\`a di
Verona, strada le Grazie 15, 
Verona 37134, Italy. email: giandomenico.orlandi@univr.it}
}

\date{}

\maketitle

\begin{abstract}
We study various properties of closed relativistic strings. In particular,
we characterize their closure under uniform convergence, 
extending a previous result
by Y. Brenier on graph-like unbounded strings, 
and we discuss some related examples. Then we
study the collapsing profile of convex planar strings which 
start with zero initial
velocity, and we obtain 
a result analogous to the well-known theorem of
Gage and Hamilton for the curvature flow of plane curves.
We conclude the paper with the discussion of an example
of weak Lipschitz evolution starting 
from the square in the plane.
\end{abstract}

\noindent
{\bf keywords:} String-theory, minimal surfaces,
Minkowski space, geometric evolutions.

\section{Introduction}
Whereas string-theory in flat Minkowski space, as viewed by physicists, is 
thought to be completely understood on the classical non-interacting level,
some of its aspects are still 
open problems from the mathematical point of view. 
The subject of this paper is the analysis of closed strings, 
which correspond to time-like minimal surfaces, 
in the $(1+n)$-dimensional flat Minkowski space.
We recall (see for instance \cite[Chapter 6]{Zw:04})
 that the Minkowski area $\area(\spacetimeimm)$ of a time-like map
$\spacetimeimm
: [0,T] \times
[0,\rightextremum
] \to \R^{1+n}$ of class $\C^1$ is given by 
\begin{equation}\label{areafunctGamma}
\area(\spacetimeimm) = \int_{[0,T]\times [0,\rightextremum]} 
\sqrt{
\langle \spacetimeimm_\paramtemp, \spacetimeimm_\paramspace\rangle_m^2 
- \langle\spacetimeimm_\paramtemp,\spacetimeimm_\paramtemp\rangle_m \langle 
\spacetimeimm_\paramspace,\spacetimeimm_\paramspace\rangle_m
} 
~d\paramtemp d\paramspace,
\end{equation}
where $\langle \cdot, \cdot \rangle_m$ denotes 
 the Minkowskian scalar product in $\R^{1+n}$ associated with the 
metric tensor  ${\rm diag}(-1,+1,\dots,+1)$.
In the sequel we always assume  
$\spacetimeimm$ to be of the form
\begin{equation}\label{gammagamma}
\spacetimeimm(t,\paramspace) := (t,\spaceimm(t,\paramspace)), \qquad (t,\paramspace)
\in [0,T] \times [0,\rightextremum], 
\end{equation}
and that 
$\spaceimm(t,\cdot)$ is closed.
It is well known (see for instance \cite{ViSh:94, Zw:04})
that in a particular parametrization (and assuming that
all quantities are sufficiently smooth) critical points
of $\area$ can be described by

\begin{equation}\label{syst:ondeintro}
\left\{
\begin{aligned}
& \spaceimm_{tt} =  \spaceimm_{xx},
\\
& \langle \spaceimm_t, \spaceimm_x\rangle=0,
\\
& \vert \spaceimm_t\vert^2 + \vert \spaceimm_x\vert^2 =1,
\end{aligned}
\right.
\end{equation}
see also Section \ref{sec:not} for the details. 
Critical points of $\area$ 
have been considered by Born and Infeld \cite{BoIn:34} (in the case of graphs), 
and analyzed later on by many authors. 
A large number of explicit solutions to
\eqref{syst:ondeintro}, possibly with 
singularities in the image of the parametrization (such as cusps, for instance,
where the above regularity condition fails)
is known, see for instance \cite[Chapter 4]{An:03},
\cite{ViSh:94},\cite{Ho:2}.

The nonlinear constraint in \eqref{syst:ondeintro} 
is not closed 
under uniform 
convergence. Indeed, many examples in the physical literature 
\cite{Ca:95, ViSh:94, Ne:90, RN:00}
show that the limit of a convergent sequence of relativistic
strings is not, in general, a relativistic string, 
thus leading to the concept of wiggly string.
A natural question is then to characterize the closure of relativistic 
strings. This issue is discussed in the physical literature (for instance in
\cite{ViSh:94}) and, for the case of strings which are entire graphs,
an answer was provided by Y. Brenier \cite{Br:05}.
We obtain an analogous result for the case of closed strings
(see Theorem \ref{th:main}). Roughly speaking, the nonlinear constraint 
is convexified (compare \eqref{ab} and \eqref{convexhull}), and limit 
solutions have in general only Lipschitz regularity. 
Then, motivated by an example
described by Neu in \cite{Ne:90} and by Theorem \ref{th:main},  
in Section \ref{sec:examples} we discuss 
various examples. In particular, 
and as already observed in \cite{Ne:90} when $n=2$ (see also \cite{Ho:2}), 
we show how
additional small oscillations superimposed on the initial datum
can prevent the limit solution to collapse to a point
(see Examples \ref{neuoscillating} and \ref{exa:giandomenico}).

Mathematical questions related to \eqref{syst:ondeintro}
also include 
the qualitative properties  of solutions for special initial data,
and their asymptotic shape near a singularity time, for instance
near a collapse. This latter problem is, in turn, intimately
related to the existence of 
weak global solutions, to be defined also 
after the onset of a singularity. 
In this paper we
begin a preliminary discussion on this subject. 
More precisely, in Section \ref{sec:convex}  
we address the study of the convexity preserving 
properties of the solutions of \eqref{syst:ondeintro}, when $n=2$, 
and their asymptotic profile near a collapsing time.
In Proposition \ref{proconvex} we show that 
a relativistic string which is smooth  and convex and has
zero initial velocity, remains convex for subsequent times, 
and shrinks to a point while its shape approaches a round
circle. This result is analogous to the one 
proven by Gage and Hamilton in \cite{GaHa:86} for curvature flow
of plane curves, and the one proven in \cite{KLW}
for the hyperbolic curvature flow (non relativistic case). 
However, differently from the parabolic case (see \cite{GaHa:86, Gr:87}), 
here the collapsing
singularity is nongeneric, 
the generic singularity being the formation of a cusp, 
as discussed in \cite{An:03, EH}. 
Adopting as a definition of weak solution the one given by 
D'Alembert formula for the linear wave system in \eqref{syst:ondeintro},
it follows that after the collapse the solution restarts,
and the motion is continued in a periodic way. 
This is in accordance with the conservative
character of the wave system in \eqref{syst:ondeintro}. 

D'Alembert formula can still
provide a possible definition
of weak solutions for Lipschitz immersions.
In  Example \ref{ex:square} we study the 
solution
corresponding to a homotetically shrinking square. 
In this case it turns out that the conservation law
\eqref{conserv_energy_integral_form} below
is valid only in special interval of times.
The same example shows that, in contrast with the case of
smooth strings,  for Lipschitz strings 
the collapsing profile is not necessarily
circular.

\section{Notation and preliminary observations}\label{sec:not}

For $n\geq 2$ we denote by $\R^{1+n}$ the $(1+n)$-dimensional
Minkowski space, which is endowed with the metric
tensor
 ${\rm diag}(-1,+1,\dots,+1)$.
We indicate by $\langle \cdot, \cdot \rangle$ and $\vert \hskip0.1truecm
\cdot \hskip0.1truecm\vert$
the euclidean scalar product and norm in $\R^n$, respectively.
Given $T>0$ and $L>0$, 
the Minkowski area $\area(\spacetimeimm)$ of 
a time-like map
$\spacetimeimm
: [0,T] \times
[0,\rightextremum
] \to \R^{1+n}$ of class $\C^1$ is defined in 
\eqref{areafunctGamma},
where 
$\spacetimeimm  = \spacetimeimm(\paramtemp,\paramspace)$,
$\spacetimeimm_\paramtemp := \partial_\paramtemp \spacetimeimm$ and
$\spacetimeimm_\paramspace := \partial_\paramspace \spacetimeimm$.
Note that \eqref{areafunctGamma} is well defined if $\spacetimeimm$
is only Lipschitz continuous.

\subsection{Assumptions on $\spaceimm$}\label{sub:theareafunct}
As already said in the Introduction, we will assume 
that 
$\spacetimeimm$ has the form \eqref{gammagamma},
where 
$\spaceimm \in \C^1([0,T] \times [0,\rightextremum];\Rn)$ satisfies
the $L$-periodicity conditions
\begin{equation}\label{closedstrings}
\spaceimm(\cdot,0)=\spaceimm(\cdot,L),
\qquad 
\spaceimm_x(\cdot,0)=\spaceimm_x(\cdot,L).
\end{equation}
When necessary, the map $\spaceimm$ will be periodically extended with 
respect to $x$ on the whole of $[0,T] \times \R$; 
we still denote by $\spaceimm\in \C^1([0,T] \times \R)$ such an extension.

\begin{Definition}\label{def:regular}
We say that $\spaceimm$ is regular if 
$\spaceimm_\paramspace(t,\paramspace)\neq 0$ for any
$(t,\paramspace) \in [0,T] \times [0,\rightextremum]$.
\end{Definition}

Let $\spaceimm \in \C^1([0,T]\times [0,L]; \Rn)$ be regular; 
if there exist a bounded closed interval $I \subset \R$
and a map ${\mathrm r} \in \C^1([0,T] \times 
I; [0,\rightextremum])$ such that
${\mathrm r}(t,\cdot)$ is strictly monotone, then 
the map 
 $(t,\sigma) \in [0,T] \times I \to
\spaceimm(t, {\mathrm r}(t,\sigma))$ is said a reparametrization of 
$\spaceimm$.

The normal velocity vector 
is given by
$\spaceimm_t^\perp$,
where $^\perp$ 
denotes the orthogonal projection onto the 
normal space, so that 
\begin{equation}\label{eq:normvelv}
\spaceimm_t^\perp
 = \spaceimm_t - \langle \spaceimm_t, \tangvect
\rangle \tangvect.
\end{equation}
\begin{Definition}
Let $\spaceimm \in \C^1([0,T] \times [0,\rightextremum];\Rn)$. 
We say that 
$\spaceimm$ is strictly admissible if
\begin{equation}\label{strictlytimelike}
\vert \spaceimm_t^\perp \vert^2 <1
\qquad {\rm in}~ [0,T] \times [0,\rightextremum].
\end{equation}
\end{Definition}
\subsection{The lagrangian $\mathcal L$}

Under assumptions \eqref{gammagamma} and \eqref{strictlytimelike} we 
have 
$$
\sqrt{
\langle \spacetimeimm_\paramtemp, \spacetimeimm_\paramspace\rangle_m^2 
- \langle\spacetimeimm_\paramtemp,\spacetimeimm_\paramtemp\rangle_m \langle 
\spacetimeimm_\paramspace,\spacetimeimm_\paramspace\rangle_m}
=
\sqrt{\langle \spaceimm_t,\spaceimm_x\rangle^2 + \vert \spaceimm_x\vert^2
(1-\vert \spaceimm_t\vert^2)},
$$
and
\begin{equation}\label{areafunct}
\begin{aligned}
\area(\spacetimeimm)  
= &  
\int_{[0,T]\times [0,\rightextremum]} \lagdens(\spaceimm_t,\spaceimm_x)~dtdx.
\end{aligned}
\end{equation}
Here the function $\lagdens : {\rm dom}(\lagdens) 
= \{(\xi,\eta) \in \R^n \times \R^n:
~ \langle \xi,\eta\rangle^2 \geq \vert \eta\vert^2
(\vert \xi\vert^2-1)
\} \to [0,+\infty)$
is defined as 
$$
\lagdens(\xi,\eta) :=  \sqrt{\langle \xi,\eta\rangle^2 + 
\vert \eta\vert^2 (1-\vert \xi\vert^2)
}, \qquad  \ \ 
(\xi,\eta) \in {\rm dom}(\lagdens).
$$
Observe that $(\xi,\eta) \in {\rm dom}(\lagdens)$ implies 
$(\xi,\alpha\eta)\in {\rm dom}(\lagdens)$ for any $\alpha
\in \R$, and 
\begin{equation}\label{posomo}
\lagdens(\xi, \alpha \eta) = \vert \alpha \vert \lagdens(\xi, \eta), \qquad 
\alpha \in \R, ~ (\xi,\eta) \in {\rm dom}(\lagdens).
\end{equation}
Hence if $\widetilde \spaceimm$ is a reparametrization of 
the regular curve $\spaceimm$ then the righe hand side of \eqref{areafunct}
remains unchanged.

Note also that
\begin{equation}\label{lag:constr}
(\xi,\eta) \in {\rm dom}(\lagdens), ~ \langle \xi,\eta \rangle =0, \quad
\vert \xi\vert^2 + \vert\eta\vert^2 = 1 
\qquad
\Rightarrow 
\qquad
\lagdens(\xi,\eta) = \vert \eta\vert^2.
\end{equation}
\begin{Definition}
Let $I\subset \R$ be a bounded closed interval,
and let 
$\spaceimm \in
 \C^1([0,T]\times I;\R^n)$ be a regular map.
We say that 
$\spaceimm$ is parametrized orthogonally  if
\begin{equation}\label{tang_vect_mutually_orthogonal}
\langle \spaceimm_t,\spaceimm_x\rangle=0 
\qquad {\rm in}~ [0,T] \times I.
\end{equation}
\end{Definition}
\begin{Remark}\label{rem:ambrosio}\rm
Any regular map $\spaceimm \in \C^1([0,T]\times 
[0,L]; \Rn)$ can be parametrized orthogonally
(see for instance 
\cite[Theorem 8]{Am:98}) 
in $[0,T] \times [0,L]$.
Indeed,
it is enough to consider the map $(t, x)\in [0,T] \times [0,L] \to  
\spaceimm(t,{\mathrm r}(t,x))$, where $\mathrm r
\in \C^1([0,T] \times [0,L]; [0,L])$
satisfies the linear transport equation
$\mathrm r_t = - \frac{\langle \spaceimm_t,
\spaceimm_\paramspace\rangle}{\vert \spaceimm_\paramspace\vert^2}
\mathrm r_\paramspace$.
It follows that the parametrization
becomes unique once we fix ${\mathrm r}(0,\cdot)$.
\end{Remark}

\smallskip

{\it Notation}:
in what follows we use the symbol $E$ with the following
meaning. 
Let $\spaceimm
 \in \C^1([0,T]\times [0,L])$
be a regular strictly admissible map. 
First we reparametrize
$\spaceimm(0,\cdot)$ 
on the interval $[0,E]$ in such a way that $\vert \spaceimm_x(0,\cdot)
\vert^2 = 
1 - \vert \spaceimm_t(0,\cdot)\vert^2$.
Next, recalling Remark \ref{rem:ambrosio}, we further 
uniquely
reparametrize orthogonally
the map $\spaceimm$ in the 
parameters space $[0,T] \times [0,E]$. We therefore achieve, 
at the same time, the two conditions 
\begin{equation}\label{constr:norminidat}
\vert \spaceimm_t(0,x)\vert^2 + 
\vert \spaceimm_x(0,x)\vert^2 =1, \qquad x \in [0,E],
\end{equation}
\begin{equation}\label{ali_delarr}
\langle \spaceimm_t(t,x), \spaceimm_x(t,x) \rangle =0, 
\qquad (t,x) \in [0,T] \times [0,E].
\end{equation}
%
\subsection{First variation of $\area$}\label{sed:firstvar}
We recall that $\spacetimeimm \in \mathcal C^1([0,T] \times [0,L]; \R^{1+n})$ 
is called a critical point of $\area$ if 
$$
\frac{d}{d\lambda} \area(\spacetimeimm + \lambda \Phi)_{\vert \lambda=0} =0,
\qquad 
\Phi \in \C^1_s\left([0,T] \times [0,L]; \R^{1+n}\right),
$$
where, for $m \in \mathbb N$, 
$\Phi \in \C^1_s\left([0,T] \times [0,L]; \R^{m}\right)$ means that
$\Phi \in \C^1\left([0,T] \times [0,L]; \R^{m}\right)$ and 
$\Phi$ has
 compact
support in $(0,T) \times [0,L]$.

The first variation of $\area$ is a classical computation 
(see for instance \cite[Section 6.5]{Zw:04}).

\begin{Lemma}\label{lem:firstvar}
Let $\spacetimeimm \in \C^1([0,T] \times [0,\rightextremum]; \R^{1+n})$ be a
critical point of $\area$ of the form \eqref{gammagamma},
with $\spaceimm$ satisfying the periodicity condition 
\eqref{closedstrings},  regular and strictly admissible.
Then 
\begin{equation}\label{eq:scalar}
\int_{[0,T] \times [0,\rightextremum]}
\frac{1}{
\lagdens(\spaceimm_t,\spaceimm_\paramspace)}
\Big(
\vert \spaceimm_\paramspace\vert^2
\psi_t
-
\langle 
\spaceimm_t,
\spaceimm_\paramspace \rangle
\psi_\paramspace
\Big)~dtdx =0,
\qquad
\psi \in 
\C^\infty_s([0,T] \times [0,\rightextremum]),
\end{equation}
\begin{equation}\label{syst_distributional}
\begin{aligned}
& \int_{[0,T]\times [0,\rightextremum]}
\frac{1}{\lagdens(\spaceimm_t,
\spaceimm_\paramspace)}
\Big[
\langle 
\spaceimm_t,
\spaceimm_\paramspace
\rangle
\Big(
\langle 
\spaceimm_\paramspace 
,\phi_t \rangle
+ 
\langle 
\spaceimm_t 
, \phi_\paramspace
\rangle 
\Big)
\\
& \qquad \qquad -
(\vert \spaceimm_t\vert^2-1)
\langle
\spaceimm_\paramspace,
\phi_\paramspace \rangle
- \vert \spaceimm_\paramspace\vert^2
\langle
\spaceimm_t 
, \phi_t \rangle\Big] ~dtdx=0, 
\end{aligned}
\quad \quad \qquad \phi \in \C^\infty_s([0,T] \times [0,\rightextremum]; \Rn).
\end{equation}
If 
$\spaceimm \in \C^2([0,T] \times [0,\rightextremum];\Rn)$ then 
\begin{equation}\label{eq:scalarepuntuale}
- \left(\frac{\vert \spaceimm_\paramspace \vert^2}{\lagdens(
\spaceimm_t, \spaceimm_\paramspace)}
\right)_t + \left(
\frac{\langle
\spaceimm_t, \spaceimm_\paramspace
\rangle}{\lagdens(\spaceimm_t, \spaceimm_\paramspace)}\right)_\paramspace =0,
\end{equation}

\begin{equation}\label{eq:EL}
\frac{\vert \spaceimm_\paramspace\vert^2 \spaceimm_{tt} 
+
\left(
\vert \spaceimm_t\vert^2-1\right) \spaceimm_{\paramspace\paramspace}
- 2 \langle 
\spaceimm_t,
\spaceimm_\paramspace
\rangle 
\spaceimm_{\paramspace t}
}
{\lagdens(\spaceimm_t,\spaceimm_\paramspace)}
+  \left[
\left(
\frac{\vert \spaceimm_t\vert^2-1}{
\lagdens(\spaceimm_t,\spaceimm_\paramspace)}
\right)_\paramspace
- 
\left(\frac{\langle 
\spaceimm_t,
\spaceimm_\paramspace
\rangle
}{\lagdens(\spaceimm_t,
\spaceimm_\paramspace)}\right)_t~\right] 
\spaceimm_\paramspace = 0
\end{equation}
in $[0,T] \times [0,\rightextremum]$.
\end{Lemma}
\begin{proof}
Let 
$\phi \in \C_s^\infty([0,T] \times 
[0,\rightextremum];\Rn)$.
For $\lambda \in \R$ and $\vert \lambda\vert$ small
enough we have that $\spaceimm + \lambda \phi$ is regular
and strictly admissible. 
Then, being $\spacetimeimm$ critical for $\area$, and taking
$\psi \in \C_s^\infty([0,T] \times [0,\rightextremum])$, we have
\begin{equation*}
\begin{aligned}
& 0
= \frac{d}{d\lambda}
\area\left(\spacetimeimm + \lambda (\psi,\phi)\right)_{\vert \lambda=0} 
\\
= & \int_{[0,T]\times [0,\rightextremum]}
\frac{d}{d\lambda}
\Bigg(
\left\langle
(1 + \lambda \psi_t, \spaceimm_t + \lambda \phi_t), 
(\lambda \psi_x, \spaceimm_x+ \lambda \phi_x)\right\rangle_m^2 
\\
&
-\left\langle 
(1+\lambda \psi_t, \spaceimm_t + \lambda \phi_t),
(1+\lambda \psi_t, \spaceimm_t + \lambda \phi_t)
\right\rangle_m
\left\langle
(\lambda \psi_x, \spaceimm_x + \lambda \phi_x),
(\lambda \psi_x, \spaceimm_x + \lambda \phi_x)
\right\rangle_m
\Bigg)^{1/2}_{\vert \lambda=0} 
~ dtdx
\\
= &  
\int_{[0,T]\times [0,\rightextremum]} \frac{
\left(
\langle 
\spaceimm_t,
\spaceimm_\paramspace
\rangle 
(\langle \spaceimm_\paramspace, \phi_t\rangle 
+ \langle 
 \spaceimm_t,
\phi_\paramspace
\rangle
- \psi_\paramspace)
- \langle \spaceimm_\paramspace, \phi_\paramspace\rangle  
(\vert \spaceimm_t\vert^2-1)
- \vert \spaceimm_\paramspace\vert^2 (
\langle \spaceimm_t, \phi_t\rangle  
- \psi_t)\right)}
{
\lagdens(\spaceimm_t,\spaceimm_\paramspace)
}\, dtd\paramspace,
\end{aligned}
\end{equation*}
and 
\eqref{eq:scalar} and
\eqref{syst_distributional} 
immediately follow.

Assume now that $\spaceimm \in \C^2([0,T] \times [0,\rightextremum]; \Rn)$. 
Then \eqref{eq:scalarepuntuale} follows from \eqref{eq:scalar} by
recalling that $\psi$ has compact support in $(0,T) \times [0,L]$, and 
integrating by parts. 
Integrating by parts in \eqref{syst_distributional} and using
\eqref{eq:scalarepuntuale}, 
it follows
\begin{equation*}
\begin{aligned}
& 
- 
\left(\frac{\langle 
\spaceimm_t,
\spaceimm_\paramspace
\rangle
}{\lagdens(\spaceimm_t,
\spaceimm_\paramspace
)}\right)_t
\spaceimm_\paramspace
- 2
\frac{\langle 
\spaceimm_t,
\spaceimm_\paramspace
\rangle
}{
\lagdens(\spaceimm_t,\spaceimm_\paramspace)} \spaceimm_{\paramspace t}
+ 
\left(
\frac{\vert \spaceimm_t\vert^2-1}{
\lagdens(\spaceimm_t,\spaceimm_\paramspace)}
\right)_\paramspace \spaceimm_\paramspace 
+ 
\frac{\vert \spaceimm_t\vert^2-1}{
\lagdens(\spaceimm_t,\spaceimm_\paramspace)} \spaceimm_{\paramspace\paramspace} 
+ 
\frac{\vert \spaceimm_\paramspace\vert^2}{
\lagdens(\spaceimm_t,\spaceimm_\paramspace)} \spaceimm_{tt} =0,
\end{aligned}
\end{equation*}
which is \eqref{eq:EL}.
\end{proof}
Note that by 
the positive one-homogeneity of $\lagdens(\xi,\cdot)$
in \eqref{posomo} it follows that 
\eqref{eq:scalar},
\eqref{syst_distributional}, \eqref{eq:scalarepuntuale}
and \eqref{eq:EL} are invariant under reparametrizations of 
$\spaceimm$ with respect to $\paramspace$. 

\begin{Remark}\label{conservation_law_integral_form}\rm 
Under the assumptions of Lemma \ref{lem:firstvar}, 
integrating \eqref{eq:scalarepuntuale} on $[0,\rightextremum]$ 
one obtains the conservation law 
\begin{equation}\label{conserv_energy_integral_form}
\frac{d}{dt} 
\int_{[0,\rightextremum]}
\frac{\vert \spaceimm_\paramspace
\vert^2}{\lagdens(\spaceimm_t,\spaceimm_\paramspace)} ~d\paramspace = 0,
\qquad t \in (0,T).
\end{equation}
This conservation law can be equivalently written on the image
$\spaceimm(t, [0,L])$ as follows:
\begin{equation}\label{cons:image}
\frac{d}{dt} 
\int_{\spaceimm(t,[0,\rightextremum])}
\frac{\theta(t,\paramspace)}{\sqrt{1-\vert \normvel(t,\cdot)\vert^2}} ~d\mathcal H^1
=0,
\end{equation}
where, given $t \in [0,T]$,  $\theta(t,x)$ is the cardinality of the set 
$\spaceimm^{-1}(t,\spaceimm(t,\paramspace))$
(in particular, $\theta(t,x)=1$ if $\spaceimm(t,\cdot)$ is an embedding),
$v := \spaceimm_t^\perp$, 
and 
$\mathcal H^1$ is the one-dimensional Hausdorff measure in $\R^n$.
 Indeed
$$
\begin{aligned}
\int_{[0,\rightextremum]}
\frac{\vert \spaceimm_\paramspace
\vert^2}{\lagdens(\spaceimm_t,\spaceimm_\paramspace)} ~d\paramspace 
= &
\int_{[0,\rightextremum]}
\frac{
\vert \spaceimm_\paramspace
\vert}{
       \sqrt{
\langle \spaceimm_t,\frac{\spaceimm_\paramspace}{\vert 
\spaceimm_\paramspace\vert}\rangle^2
           + 1 - \vert \spaceimm_t\vert^2}}
~d\paramspace 
= 
\int_{[0,\rightextremum]}
\frac{
\vert \spaceimm_\paramspace
\vert}{
\sqrt{
1 - \vert \spaceimm_t^\perp\vert^2}} ~d\paramspace 
\\
= &
\int_{\spaceimm(t,[0,\rightextremum])}
\frac{\theta(t,\paramspace)}{\sqrt{1-\vert \normvel(t,\cdot)\vert^2}} ~d\mathcal H^1
=0,
\end{aligned}
$$
where the last equality follows from the area formula \cite{AmFuPa:00}.
\end{Remark}
\begin{Corollary}\label{cor:onde}
Assume that $\spaceimm\in \C^1([0,T]
\times [0,\rightextremum];\R^n)$ is regular, strictly 
admissible,
and satisfies \eqref{eq:scalar} and 
\eqref{syst_distributional}. Define
$$
\rho(x) := \frac{|\spaceimm_x(0,x)|}{\sqrt{1- \vert \spaceimm_t(0,x)\vert^2}},
\qquad 
x \in [0,\rightextremum].
$$
If 
$\spaceimm$ is parametrized orthogonally  
then 
\begin{itemize}
\item[(i)] the conservation law
\eqref{conserv_energy_integral_form} 
strengthen into the pointwise
conservation law 
\begin{equation}\label{eqphi}
\frac{|\spaceimm_x(t,x)|}{\sqrt{1- \vert \spaceimm_t(t,x)\vert^2}} = 
\rho(x), 
\qquad 
(t,x) \in [0,T] \times [0,\rightextremum];
\end{equation}
\item[(ii)] the condition
 \eqref{syst_distributional} becomes
$$
\int_{[0,T]\times [0,\rightextremum]} \langle \spaceimm_t \rho, 
\phi_t\rangle~dtdx 
= \int_{[0,T]\times [0,\rightextremum]} \langle \frac{\spaceimm_x}{\rho},
\phi_x \rangle
~dtdx;
$$
\item[(iii)]
if we reparametrize $\spaceimm(0,\cdot)$ on the
interval $[0, \energy]$ so that 
$\rho$ is constantly equal to $1$, that is if \eqref{constr:norminidat}
holds,
then 
\begin{equation}\label{secondovincolo}
\vert \spaceimm_t \vert^2 + \vert \spaceimm_x \vert^2=1
\qquad {\rm ~in}~ [0,T]\times [0,\energy],
\end{equation}
and $\spaceimm$ becomes a $\mathcal C^1$ 
distributional solution of the wave linear system
\begin{equation}\label{wavelinearsystem}
\spaceimm_{tt} = \spaceimm_{xx}
\qquad {\rm in}~ [0,T] \times [0,\energy].
\end{equation}
\end{itemize}
\end{Corollary}
\begin{proof}
Using the orthogonality condition 
\eqref{tang_vect_mutually_orthogonal}
we have 
\begin{equation}\label{firstcommentjens}
\lagdens(\spaceimm_t,\spaceimm_x) = \vert \spaceimm_x\vert \sqrt{1 - \vert
\spaceimm_t\vert^2},
\end{equation}
and
equation 
\eqref{eq:scalar} reduces to
\[
\int_{[0,T] \times [0,\rightextremum]}
\frac{|\spaceimm_x|}{\sqrt{1- \vert \spaceimm_t\vert^2}}
~ \psi_t~dtdx = 0, \qquad \psi \in \C^\infty([0,T] \times [0,L]),
\]
which implies \eqref{eqphi}. 

{}From \eqref{syst_distributional} and the fact that 
the parametrization of $\spaceimm$ is orthogonal, we obtain
\begin{equation}\label{C1normal}
\int_{[0,T]\times [0,\rightextremum]}
\left(
\langle \frac{\spaceimm_x (1-\vert \spaceimm_t\vert^2)}{
\lagdens(\spaceimm_t,\spaceimm_x)}
,\phi_x \rangle
-
\langle \frac{\spaceimm_t \vert \spaceimm_x\vert^2}{
\lagdens(\spaceimm_t,\spaceimm_x)
},
\phi_t\rangle \right)~dtdx =0.
\end{equation}
Using \eqref{firstcommentjens} it then follows
\begin{equation*}
\int_{[0,T]\times [0,\rightextremum]}
\left(
\langle
\frac{\spaceimm_x \sqrt{1-\vert\spaceimm_t\vert^2}}{\vert \spaceimm_x\vert}
,\phi_x \rangle
-
\langle
\frac{\spaceimm_t \vert \spaceimm_x\vert}{
\sqrt{1-\vert \spaceimm_t\vert^2})
}
, \phi_t\rangle\right)~dtdx =0,
\end{equation*}
that is 
\begin{equation*}
\int_{[0,T]\times [0,\rightextremum]}
\left(
\langle
\frac{\spaceimm_x}{\rho}
,\phi_x \rangle
-
\langle
\spaceimm_t \rho, \phi_t\rangle\right)~dtdx =0,
\end{equation*}
which is (ii).

Eventually, assertion (iii) follows directly from (i) and (ii).
\end{proof}
\begin{Remark}\label{rem:constr_manteined}\rm
We point out that
Corollary \ref{cor:onde} (iii) shows that
 if the constraint $\vert \spaceimm_t\vert^2
+\vert \spaceimm_x\vert^2=1$ is valid at the initial time $t=0$,
then it remains valid at subsequent times.
\end{Remark}

A number of solutions of \eqref{tang_vect_mutually_orthogonal}, 
\eqref{secondovincolo}, \eqref{wavelinearsystem} are known,
see for instance \cite[Section 6.2.4]{ViSh:94}, \cite[Chapter 4]{An:03},
\cite{GiIs:04}, the simplest one being probably the following
\cite{Ne:90}. 
Let $n=2$, $R>0$  and $a(s) = b(s) 
:= R (\cos\frac{s}{R}, \sin\frac{s}{R})$ for any $s \in \R$.
The solution to \eqref{wavelinearsystem} becomes 
$$
\spaceimm(t,x) = R \left(
\cos \frac{x}{R}, \sin \frac{x}{R}\right) \cos\frac{t}{R}, \qquad 
(t,x) \in \left(-R\pi/2, R \pi/2\right) \times [0,\energy],
$$
with $\energy = 2\pi R$.
Note that 
at the singular times $t = \pm \energy/4$, 
the condition $\spaceimm_x(t,\cdot) \neq 0$ 
is not satisfied, and  $\spaceimm(t, [0,\energy])$ reduces to a point.

\subsection{Representation of the solutions and a concept
of weak solution}\label{subsect:rep}
Let $\spacetimeimm\in \C^1([0,T]\times [0,E]; \R^{1+n})$ 
(resp. $\spacetimeimm\in \C^2([0,T]\times [0,E]; \R^{1+n})$)
be a critical point of $\area$ of the form \eqref{gammagamma}, where
$\spaceimm \in \C^1([0,T]\times [0,\rightextremum]; \Rn)$ 
(resp. $\spaceimm \in \C^2([0,T] \times [0,
\rightextremum]; \Rn)$)
is  strictly admissible and regular.
We have seen that 
there exists an orthogonal 
parametrization
of $\spaceimm$ satisfying
\eqref{constr:norminidat},
hence by Corollary \ref{cor:onde} (iii) we have that 
$\spaceimm$
becomes a distributional
(resp. classical) solution to \eqref{wavelinearsystem}. 
Hence 
there exist
$\energy$-periodic maps
$a, b \in \C^1(\R;\R^n)$ (resp. $\C^2(\R; \Rn)$) such that 
\begin{equation}\label{eq:representgamma}
\spaceimm(t,x) = 
\frac{1}{2} 
\left[ a(x+t) + b(x-t)
\right], \qquad (t,x) \in [0,T] \times [0,
\energy
],
\end{equation}
\begin{equation}\label{ab}
\vert a'\vert = \vert b'\vert = 1 \qquad {\rm in}~ \R.
\end{equation}
Note that $\spaceimm_t(0,\cdot)=0$ if and only if there exists
$w \in \Rn$ such that $a = b + w$.

\smallskip

\begin{Remark}\label{rem:mettitogli}\rm 
Since $a$ and $b$ are defined on 
the whole of $\R$, the right hand 
side of \eqref{eq:representgamma} can 
be considered as the definition of the 
 map
$\spaceimm$ on the left hand side also for $(t,x) \in (\R \setminus [0,T])
\times [0,\energy]$. 
Namely, the right hand side of \eqref{eq:representgamma} 
provides a global in time $\mathcal C^1$  
(resp. $\mathcal C^2$) weak solution, denoted
by $\weaksol$
to \eqref{tang_vect_mutually_orthogonal}, \eqref{secondovincolo} and 
 \eqref{wavelinearsystem} defined for
$(t,\paramspace) \in \R \times [0,E]$. In general 
it may happen that $\weaksol_x(\overline t,\overline x)=0$ for some 
$(\overline t, \overline x) \in (\R\setminus[0,T]) \times [0,\energy]$, since
\eqref{ab}
does not prevent that $a'(\overline x+ \overline t)=-b'(\overline 
x- \overline t)$. Hence
  singularities
in the image $\weaksol(\overline t,
[0,E])$ (such as cusps, for instance) are in 
general expected,
and may possibly persist in time (see Remark \ref{rem:persist} below). 
We point out that such a weak solution 
could not coincide with the weak solution proposed in \cite{BeNoOr:09} 
when 
singularities are present.  
Another notion of weak solution 
to the lorentzian minimal surface equation in the case of 
graphs has been proposed in \cite{Br:05}. 
\end{Remark}

We conclude
this section by observing that 
the time-slices $\gamma(t,\cdot)$ of a surface which
is critical for $\area$
 satisfy
the geometric equation
\begin{equation}\label{eqgeom}
{\mathrm a} = (1- \vert v\vert^2)\ \kappa.
\end{equation}
Here, 
if $\spaceimm \in \C^2([0,T]\times [0,\rightextremum]; \R^n)$ is regular,
$v =\spaceimm_t^\perp$ denotes the normal velocity vector, $\kappa$ denotes 
 the curvature vector 
and ${\mathrm a}$ 
the normal
acceleration vector,
respectively given\footnote{
When $\spaceimm$ is an embedding, if we set
$\Gamma(t) := \spaceimm(t,[0,L])$, 
then ${\mathrm a}
= (\grad \eta_{tt})^\perp$ on $\Gamma(t)$, 
where $\eta(t,z) := {\rm dist}(z, 
\Gamma(t))^2/2$ for $(t,z) \in [0,T] \times \Rn$. In the case 
$n=2$ it holds $v=-d_{t}\nabla d$ and
${\mathrm a} = -d_{tt} \nabla d$, 
where $d$ is the signed distance from $\Gamma(t)$.} 
by 
\begin{equation}\label{eq:kappava}
\kappa = \frac{\spaceimm_{\paramspace\paramspace}^\perp}{\vert 
\spaceimm_\paramspace\vert^2}\,,
\qquad 
{\mathrm a} = \left( 
v_{t} -
\langle \spaceimm_t, \frac{\spaceimm_x}{
\vert\spaceimm_x\vert}
\rangle
\frac{v_\paramspace}{\vert\spaceimm_x\vert} 
\right)^\perp = \,\spaceimm_{tt}^\perp +
\langle \spaceimm_t, \frac{\spaceimm_x}{
\vert\spaceimm_x\vert}
\rangle \left( \frac{\spaceimm_{\paramspace\paramspace}}{\vert 
\spaceimm_\paramspace\vert^2} - 2
\frac{\spaceimm_{\paramspace t}}{\vert\spaceimm_x\vert}
\right)^\perp. 
\end{equation}
To show \eqref{eqgeom}, observe that 
$$
\begin{aligned}
(\spaceimm_t^\perp)_t^{~\perp} = & 
\spaceimm_{tt}^\perp - \langle
\spaceimm_t, \tangvect\rangle \left(\tangvect\right)_t^{~\perp} 
= 
\spaceimm_{tt}^\perp - \langle
\spaceimm_t, \tangvect\rangle \frac{\spaceimm_{t\paramspace}^\perp}{\vert
\spaceimm_\paramspace\vert},
\\
(\spaceimm_t^\perp)_\paramspace^{~\perp} = & 
\spaceimm_{t\paramspace}^\perp - \langle
\spaceimm_t, \tangvect\rangle \left(
\tangvect\right)_\paramspace^{~\perp} 
= 
\spaceimm_{t\paramspace}^\perp - 
\langle
\spaceimm_t, \tangvect\rangle \frac{\spaceimm_{\paramspace\paramspace}^\perp}{\vert
\spaceimm_\paramspace\vert},
\end{aligned}
$$
so that 
\begin{equation}\label{prel_express}
{\mathrm a} 
= \spaceimm_{tt}^\perp -  2\langle \spaceimm_t, \tangvect
\rangle\, \frac{\spaceimm_{tx}^\perp}{|\spaceimm_x|} - 
\langle
\spaceimm_t, \tangvect\rangle^2 
~\frac{\spaceimm_{xx}^\perp}{\vert \spaceimm_x\vert^2},
\end{equation}
and therefore if $\spaceimm$ is parametrized
orthogonally, then 
${\mathrm a} = (\spaceimm_t^\perp)_t^\perp$.
Now, projecting both sides of 
\eqref{eq:EL} onto the normal space to $\spaceimm(t,\cdot)$
gives 
\begin{equation}\label{projELonthenormalspace}
\spaceimm_{tt}^\perp
+
\frac{
\vert \spaceimm_t\vert^2-1
}{\vert \spaceimm_\paramspace\vert^2}
\spaceimm_{\paramspace\paramspace}^\perp
- 2 \frac{\langle 
\spaceimm_t,
\spaceimm_\paramspace
\rangle}{\vert \spaceimm_\paramspace\vert^2} 
\spaceimm_{\paramspace t}^\perp
=0.
\end{equation}
Inserting \eqref{prel_express} into \eqref{projELonthenormalspace} gives
$$
{\mathrm a} = 
\frac{
1-\vert \spaceimm_t\vert^2
- \langle \spaceimm_t,\tangvect\rangle^2}{\vert \spaceimm_x\vert^2}
\spaceimm_{xx}^\perp
= \left( 1- |\spaceimm_t^\perp|^2\right)\frac{\spaceimm_{xx}^\perp}{|\spaceimm_x|^2}
= (1-\vert v\vert^2)\kappa.
$$
%

\section{Closure of solutions}\label{sec:main}
The closure result is 
 motivated by an example in \cite{Ne:90} ( see also the discussion
in \cite[Section 6.5.2]{ViSh:94}, and references therein ), 
and is similar 
to the one in \cite{Br:05}, where
maps which are graphs 
defined in the whole of $\R \times \R$ are considered.
\begin{Theorem}\label{th:main}
Let $\{\energy_k\}$ be a sequence of positive numbers
converging to
$\energy  \in [0,+\infty)$ as $k \to +\infty$.
Let 
$\{\spaceimm_k\}\subset \C^1([0,T]\times \R;\R^n)$ 
be a sequence of $E_k$-periodic regular strictly admissible 
orthogonally parametrized
maps 
\begin{equation}\label{constr:norminidatkappa}
\vert {\spaceimm_k}_t(0,x)\vert^2 + 
\vert {\spaceimm_k}_x(0,x)\vert^2 =1, \qquad x \in \R,
\end{equation}
and
solving the wave system \eqref{wavelinearsystem}. 
The following assertions hold.
\begin{itemize}
\item[(i)]  if
$\{\spaceimm_k\}$ 
converges 
to a map $\spaceimm \in {\rm Lip}([0,T]\times \R;\Rn)$ uniformly
in  $[0,T] \times \R$ as $k \to +\infty$,
then there exist $\energy$-periodic maps 
$a, b \in {\rm Lip}(\R; \Rn)$ 
with
\begin{equation}\label{convexhull}
\vert a'\vert \leq 1, \qquad
\vert b'\vert \leq 1 \qquad {\rm a.e.~in}~ \R
\end{equation}
such that 
$\spaceimm$ has the representation \eqref{eq:representgamma} in 
$[0,T] \times \R$.
\item[(ii)]
If $\spaceimm \in {\rm Lip}([0,T] \times [0,\energy];\Rn)$
can be represented as in \eqref{eq:representgamma}
where
$a, b \in {\rm Lip}(\R; \Rn)$ are
$\energy$-periodic maps
satisfying \eqref{convexhull}, 
then there  exists
a sequence 
$\{\spaceimm_k\} \subset \C^2([0,T]\times \R;\Rn)$ of 
$\energy$-periodic maps
solving 
\eqref{wavelinearsystem}, \eqref{tang_vect_mutually_orthogonal}, 
\eqref{constr:norminidat} in $[0,T] \times \R$ 
such that  
$\{\spaceimm_k\}$ converges to $\spaceimm$ uniformly 
in  $[0,T] \times [0,\energy]$.
\end{itemize}
\end{Theorem}
\begin{proof}
Let us prove (i). 
Let $a_k, b_k$, with $|a_k'|=|b_k'|=1$, be such that 
\eqref{eq:representgamma} holds with $\energy, \spaceimm,a,b$ 
replaced by $\energy_k,\spaceimm_k,a_k,b_k$, respectively. Then
assertion (i) 
follows  by recalling that, if $L> \sup_k \energy_k$,
 the set 
$\{ u\in W^{1,\infty}
(\R;\R^n):
u {\rm ~is}~L~{\rm periodic},  |u'|\le 1 ~{\rm a.e.}\}$ 
is the weak$^*$ closure 
of $\{ u\in W^{1,\infty}
(\R;\R^n):
u {\rm ~is}~L~{\rm periodic},  |u'|= 1 ~{\rm a.e.}\}$, 
and in particular it is 
closed
under the uniform convergence on the compact subsets.

\medskip

Let us prove (ii). 
Given $a,b\in {\rm Lip}(\R;\R^n)$ 
$\energy$-periodic maps satisfying
 $|a'|\le 1$ and $|b'|\le 1$ almost everywhere,
it is enough to find two $\energy$-periodic sequences 
$\{a_k\}$, $\{b_k\}\subset \C^2(\R;\R^n)$, 
with $|a_k'|=|b_k'|=1$, uniformly converging to $a,b$, respectively,
as $k \to \infty$.
It is also sufficient to prove this assertion for $a,b$ 
belonging to the dense (for the uniform convergence) 
class of piecewise linear
immersions satisfying \eqref{convexhull}, since one then 
concludes for general $a,b$ using a diagonal argument. 
We will 
show the assertion for the map $a$, the construction for $b$ being similar.
Let $a$ be an $E$-periodic
piecewise linear immersion
satisfying \eqref{convexhull}, 
so that we can identify the points $\{0\}$ and $\{ E\}$,  
and assume that there exist $m+2$ points $0=:L_0<L_1<\dots <L_{m+1}:=\energy$ 
in the interval $[0,E]$ such that 
\[
a(x)=a(L_i)+(x-L_i)c_{i+1}, \qquad x \in [L_i, L_{i+1}], \ \ 
i =0, \dots, m,
\] 
with $c_{i+1}\in\R^n$, 
$|c_{i+1}|\le 1$ for $i=0,\dots,m$. 
Choose $d_{i+1}\in\R^n$ so that 
\begin{equation}\label{cidi}
\langle d_{i+1},
c_{i+1}\rangle=0, \qquad
|d_{i+1}|^2=1-|c_{i+1}|^2, \qquad \quad
i=0,\dots,m.
\end{equation}
Fix $k\in\N$ even. For $i=0,\dots,m$ we take a partition
of $[L_i, L_{i+1}]$ into $k$ subintervals of 
equal length: precisely, i.e.,
 $j=0,\dots,k$ 
set $L_i^j:=L_i+\frac{j}{k}(L_{i+1}-L_i)$ (we write
$L_i^0=L_i$ and $L_i^k = L_{i+1}$). Define
\begin{equation}\label{def:ak}
\bar a_k(x):=a(x)+(-1)^j (x-L_i^j) d_{i+1}, \qquad
L_{i}^j\le x\le L_i^{j+1},
\end{equation}
see Figure \ref{fig:closure} (a).
 Since $k$ is even, $\bar a_k \in {\rm Lip}([0,E]; \Rn)$. Moreover 
from \eqref{cidi} it follows
 $|\bar a'_k(x)|=1$ 
for any $x\in [0,\energy]$ out of a finite set depending on $k$. Eventually, 
by construction $|\bar a_k(x)-a(x)|\le \frac{L}{k}$ 
for any $x\in [0,\energy]$, 
so that $\bar a_k\to a$ uniformly in $[0,\energy]$ as
$k \to +\infty$. Once a similar construction
for $b$ (thus leading to the definition of $\{\bar b_k\}$) is made,
let us consider  the sequence $\{\spaceimm_k\}$
of maps defined as $\bar \spaceimm_k(t,x) := 
\frac{1}{2} \left[
\bar a_k(\paramspace+t) + \bar b_k(\paramspace-t)\right]$ for any
$(t,x) \in [0,T] \times [0,E]$. 
These maps belong to 
${\rm Lip}([0,T] \times [0,\energy] ;\Rn)$, 
and must be regularized in order to avoid the presence
of corners.

\begin{figure}
\begin{center}
\includegraphics[height=1.5cm]{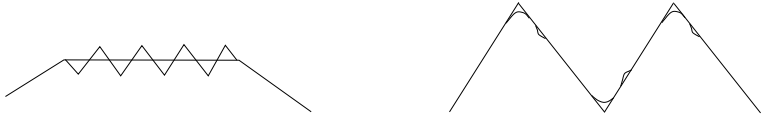}
\caption{\small (a): the construction of $\bar a_k$ defined in \eqref{def:ak} 
for a piecewise 
linear map $a$, in the (image of) the
interval $[L_i, L_{i+1}]$; the slopes are $c_{i+1}\pm d_{i+1}$.
 (b): the smoothing of the corners
in order to have $\bar a_k \in \C^2$, keeping
the length constraint satisfied.
} 
\label{fig:closure}
\end{center}
\end{figure}

\smallskip

Given $k \in \mathbb N$, $k \geq 1$, let $\ell_k \in 
\displaystyle \left(0,\min_{i=0,\dots,m} \frac{L_{i+1}-L_i}{3k}\right)$ 
and fix $\eta \in (0,\ell_k/3)$. We apply Lemma \ref{lem:spline} below 
with $\ell=\ell_k$ 
in the intervals $[L_i^j-\ell_k,L_i^j+\ell_k]$, identifying 
$\R^2$ with $\bar a_k(L_i^j)+ {\rm span}\{c_{i+1}+d_{i+1}, c_{i+1}-d_{i+1}\}$ 
if $j\neq 0$, and with $\bar a_k(L_i)+ 
{\rm span}\{c_{i+1}+d_{i+1}+c_{i}-d_{i}, c_{i+1}+d_{i+1}-c_{i}+d_{i}\}$ if $j=0$
(see Figure \ref{fig:closure} (b)).
%
In both cases set $s:=x-L_i^j$ and 
$\spaceimm_{ijk}(s):=\bar a_k(s+L^j_i)= \bar a_k(x)$. 
Let $\widetilde\spaceimm_{ijk}$ be 
the approximations of $\spaceimm_{ijk}$ obtained by 
Lemma \ref{lem:spline}.  
$s\mapsto a_k(s)$ for $0\le s\le \ell_k$, 
Then, the map 
$$
a_k(x):=
\begin{cases}
\widetilde\spaceimm_{ijk}(x-L_i^j) &\text{if }  L_i^j-\ell_k\le x\le L_i^j+\ell_k\\
\bar a_k(x) &\text{otherwise in } [0,\energy],
\end{cases}
$$
extended by $\energy$-periodicity, is of class $\C^2(\R)$, $|a_k'|=1$ and 
$\Vert a_k-\bar a_k\Vert_{L^{\infty}([0,\energy])}\le L/k$. 
\end{proof}

\begin{Lemma}\label{lem:spline} 
Let $\ell>0$, $(\tau_1,\tau_2)\in\R^2$ be a unit vector 
such that  $\tau_1,\tau_2>0$, 
and let  $\spaceimm(s):=(s\tau_1, |s|\tau_2)$ for $-\ell\le s\le \ell$. 
For any $\eta \in (0,\ell/3)$ there exists 
$\widetilde\spaceimm\in \mathcal C^2([-\ell,\ell];\R^2)$
such that $|\widetilde\spaceimm'|=1$ in $[-\ell, \ell]$,
$\widetilde\spaceimm(s)=\spaceimm(s)$ for $|s|\ge\frac{\ell}{2}$ and 
$\Vert\spaceimm-\widetilde\spaceimm\Vert_{L^\infty([-\ell,\ell])}\le \eta$.
\end{Lemma}

\begin{proof}
Consider without loss
of generality $\ell<1$, 
fix $\eta \in (0,\frac{\ell}{3})$ 
and let $0<\alpha,\beta\le \eta/2$ be two parameters to be fixed later. Define
the map $\spaceimm_{\alpha,\beta} \in \C^2([-\ell,\ell]; \R^2)$ as
$$
\spaceimm_{\alpha,\beta}(\paramgiandoa):=
\begin{cases}
 (\tau_1 \paramgiandoa, -\frac{\tau_2}{8\alpha^3}\paramgiandoa^4
+\frac{3\tau_2}{4\alpha}\paramgiandoa^2+\frac{3\tau_2\alpha}{8}) 
&\text{if } |t\paramgiandoa|\le \alpha\\
 \paramgiandoa(\tau_1,\tau_2)+\beta(\paramgiandoa-\frac{\ell}{3})^3(\frac{\ell}{2}-
\paramgiandoa)^3(-\tau_2,\tau_1) &\text{if } \frac{\ell}{3}\le \paramgiandoa
\le \frac{\ell}{2}\\
  (\tau_1 \paramgiandoa,\tau_2|\paramgiandoa|) 
&\text{otherwise in } [-\ell, \ell], 
\end{cases}
$$
see Figure \ref{fig:closure} (b). The definition in $[-\alpha,
\alpha]$ corresponds to the smoothened corners, while the definition
in $[\ell/3, \ell/2]$ corresponds to the small ``bump''
out of the corners.
 
For $\alpha$ and $\beta$ sufficiently small
$|\spaceimm_{\alpha,\beta}(\paramgiandoa)-\spaceimm(\paramgiandoa)|\le 
\eta/2$. Moreover 
$$
\int_{-\alpha}^\alpha|\spaceimm'_{\alpha,\beta}|\, d\paramgiandoa  < 
2\alpha, \qquad \int_{\ell/3}^{\ell/2}|\spaceimm'_{\alpha,\beta}|\, 
d\paramgiandoa\, >\frac{\ell}{6}\, ,
$$
and $$\int_{\ell/3}^{\ell/2}
|\spaceimm'_{\alpha,\beta}|\, d\paramgiandoa\, 
\to\frac{\ell}{6} \qquad\text{as } \beta\to 0^+.
$$
Hence there exist $\alpha$ and $\beta$ such that 
$$
\int_{-\alpha}^\alpha|\spaceimm'_{\alpha,\beta}|\, d\paramgiandoa\, 
+\, \int_{\ell/3}^{\ell/2}|\spaceimm'_{\alpha,\beta}|\, d\paramgiandoa\,=
2\alpha+\frac{\ell}{6}\, 
$$
and in particular 
\begin{equation}\label{eq:alfabeta}
\int_{-\ell/2}^{\ell/2}|\spaceimm'_{\alpha,\beta}|d\paramgiandoa=
\frac{\ell}{2} - \alpha  
+ \frac{\ell}{3} - \alpha + 2 \alpha + \frac{\ell}{6}=
\ell\, .
\end{equation} 
Let $s=s(\paramgiandoa)$ be 
the arc-length parameter for the curve $\spaceimm_{\alpha,\beta}$, and observe 
that $|s-\paramgiandoa(s)|\le 2\alpha$. Set 
$\widetilde\spaceimm(s):=\spaceimm_{\alpha,\beta}(\paramgiandoa(s))$. 
By \eqref{eq:alfabeta} 
we deduce that $\widetilde\spaceimm(s)=\spaceimm(s)$ for $|s|\ge\ell/2$. Moreover,
$$
\begin{aligned}
|\widetilde\spaceimm(s)-\spaceimm(s)| &=|\spaceimm_{\alpha,\beta}(\paramgiandoa(s))
-\spaceimm(s)|\le |\spaceimm_{\alpha,\beta}(\paramgiandoa(s))
-\spaceimm(\paramgiandoa(s))|+|\spaceimm(\paramgiandoa(s))-\spaceimm(s)|\\
&\le \frac{\eta}{2}+|\paramgiandoa(s)-s|
\le \frac{\eta}{2}+2\alpha\le \eta\, .\\
\end{aligned}
$$
\end{proof}

\begin{Remark}\label{rem:neu}\rm
Let $\spaceimm\in \C^2([0,T] \times [0,\rightextremum]; \Rn)$ be a regular
strictly admissible orthogonally parametrized map.
Then \eqref{prel_express} implies that 
\begin{equation}\label{ppp}
{\mathrm a} = \spaceimm_{tt}^\perp. 
\end{equation}
If $\spaceimm$ in addition  satisfies the wave system 
\eqref{wavelinearsystem},
so that the representation formula 
\eqref{eq:representgamma} holds, but assuming  only
$\vert a'\vert = \vert b'\vert \leq 1$ instead of \eqref{ab},
then being $\spaceimm_{tt}^\perp = \spaceimm_{xx}^\perp$, we have the
identity 
\[
{\mathrm a}  = (1-
\vert v\vert^2)\kappa - (1-\vert v\vert^2)\kappa + 
\alpha\left( \spaceimm_{xx}^\perp -\spaceimm_{tt}^\perp\right) + \spaceimm_{tt}^\perp,
\qquad \alpha \in \R.
\]
Choosing $\alpha=2(1- \vert v\vert^2)/(1-\vert v\vert^2 
+ |\spaceimm_x|^2)$ we get, using 
$\kappa \vert \spaceimm_x\vert^2 = \spaceimm_{xx}^\perp$ and \eqref{ppp}, 
\begin{equation}\label{eqsto}
\begin{aligned}
- {\mathrm a} + (1-
\vert v\vert^2)\kappa 
= & \frac{1 - \phi^2}{1 + \phi^2}\left(
{\mathrm a} + (1-\vert v\vert^2)\kappa\right),
\end{aligned}
\end{equation}
where 
$\phi := \frac{|\spaceimm_x|}{\sqrt{1 - |\spaceimm_t|^2}}$.
In analogy with the discussion in \cite[Section 5]{Ne:90},
the left-hand side of \eqref{eqsto} is the mean curvature
of the surface,
while the right-hand side 
can be interpreted as a sort of sectional curvature of the surface in the 
null direction,
multiplied by the positive factor
 $\frac{1-\phi(t,x)^2}{1+\phi(t,x)^2}$.
\end{Remark}

\section{Some examples}\label{sec:examples}
In view of Theorem \ref{th:main}, we are interested
in understanding the structure of the uniform limits 
of $\C^2$ critical points of the 
functional $\area$ of the form \eqref{gammagamma}. The following
example shows that such limits cannot satisfy, in general, any kind of partial
differential equation.
\begin{Example}\label{exa:cylinder}\rm 
Let $A\in\C^2(\R; \Rn)$
 be an $L$-periodic map
 satisfying $|A'|=1$. 
Let also 
$\eps \in (0,1)$ be such that $L/\eps\in\mathbb N$,
and define $B_\eps : \R \to \R$ as 
$$
B_\eps(s) := \eps A\left(\frac{s}{\eps}\right), \qquad s \in \R.
$$ 
Let $E = 2 L$. Define
$$
\weaksol_\eps(t,x) := \frac{1}{2} \left[
2A\left(\frac{x+t}{2}\right) + B_\eps(x-t)
\right], \qquad (t,x)\in \R\times [0,E].
$$
Then $\weaksol_\eps \in \C^2(\R\times [0,2\rightextremum]; \Rn)$,
it satisfies 
$\langle 
{{\weaksol_\eps}}_t, {{\weaksol_\eps}}_x
\rangle=0$, 
$\vert {{\weaksol_\eps}}_t\vert^2
 + \vert {{\weaksol_\eps}}_x \vert^2 =1$
 and it is a global in time solution of \eqref{wavelinearsystem}. 
The maps $\weaksol_\eps(t,x)$ converge, as $\eps\to 0^+$, 
to $A((x+t)/2)$ uniformly on the compact subsets
of $\R \times [0,\energy]$, and $A((x+t)/2)$
is a 
reparametrization of $A(x/2)$, $x\in [0,\energy]$. 
In particular,  
if $\spaceimm_0 \in \C^2([0,L]; \Rn)$ is a closed regular
curve, the curve $\spaceimm(t,x) := \spaceimm_0(x)$
(the image of which is the ``cylinder'' $\R \times \spaceimm_0([0,L])$)
is a local uniform limit 
of a sequence corresponding to $\C^2$-critical 
points of the functional $\area$.
\end{Example}

The next example should be compared with the example 
given by Neu in \cite{Ne:90}, and with the one in \cite[Section 1]{Br:05}.

\begin{Example}\label{neuoscillating}\rm 
Let $n=2$ and $a(s) :=(\cos s, \sin s)$ for any $s \in \R$.
We want to approximate uniformly the pair $(a(s),a(s))$ with 
pairs which have approximately the form $(a(s), a(s) + \frac{1}{2n} 
a(ns))$, where $n \in \mathbb N$.
Since we want to keep the constraints in \eqref{ab}, and in addition
we want to control the periods,
we need to make suitable reparametrizations. 
The conclusion of the example will be that 
there exists $\alpha >1$ (see \eqref{alfa} below) such that the map
\begin{equation}\label{theconclusion}
\spaceimm(t,x) = \frac{1}{2} \left[\alpha a\left(\frac{x+t}{\alpha}\right) 
+ a\left(\frac{x-t}{\alpha}\right)\right]
\end{equation}
 can be obtained as 
local uniform limit of (the  second components,
see \eqref{gammagamma})  a sequence of $\C^2$ critical points
of $\area$.  In particular, the presence of $\alpha>1$ 
prevents 
$\spaceimm(t,x)$ to vanish, since \eqref{theconclusion} implies
$$
\vert \spaceimm(t,x)\vert^2 = \frac{1}{4} \left(
1+\alpha^2 + 2 \alpha \cos(2t/\alpha)
\right) \geq \frac{(1-\alpha)^2}{4}>0.
$$
We begin by introducing
the smooth strictly increasing function $s_n:  \R \to \R$, having
a $2\pi$-periodic derivative, and vanishing at $0$,  
as follows: for any $x \in \R$ we set
\begin{equation*}
\begin{aligned}
s_n(x) := &
\int_0^x \left\vert a'(\sigma) + \frac{1}{2} a'(n\sigma)
\right\vert~d\sigma
\\
= & \int_0^x \sqrt{\left[
\sin 
\varmuta
 + \frac{1}{2}\sin (n\varmuta
)
\right]^2
+ \left[\cos \varmuta
 + \frac{1}{2}\cos(n\varmuta
) \right]^2}~d\varmuta
= \int_0^x
 \sqrt{\frac{5}{4} + \cos\left(n\varmuta
 -\varmuta
\right)}~
d\varmuta.
\end{aligned}
\end{equation*}
Observe that $s_n'\geq \frac{1}{2}$ everywhere. Set 
$$
\ell_n:= s_n(2\pi),
$$
and denote by $x_n: \R \to \R$  the inverse of $s_n$.
Next define
\begin{equation}\label{bepsarclength}
b_n(s) 
:= a(x_n(s)) + \frac{1}{2n} a\left(nx_n(s)\right),
\qquad s \in \R.
\end{equation}
Notice that $b_n$ is 
$\ell_n$-periodic, since 
given $k \in \mathbb N$ we have 
$s_n(2k\pi) = k \ell_n$ and 
$x_n(k\ell_n) = 2k\pi$. 
Furthermore
\begin{equation}\label{bepsuno}
\vert b_n'(s)\vert = 
\vert x_n'(s)\vert \left\vert a'(x_n(s)) + 
\frac{1}{2} a'(nx_n(s))\right\vert = 1,
\qquad s \in \R.
\end{equation}
The period $\ell_n$ is larger than 
the period of $a$, 
due to the presence of the additional
oscillations. 

Let also 
$$
a_n(s) := \frac{\ell_n}{2\pi} 
a\left(
\frac{2\pi s}{\ell_n}
\right), \qquad s \in \R.
$$
The map  $a_n$ has the same period as $b_n$
and satisfies 
\begin{equation}\label{aepsuno}
\vert a_n'\vert=1.
\end{equation}
Define
$$
\spaceimm_n(t,x) := \frac{1}{2} \left[ a_n(x+t) + b_n(x-t)
\right], \qquad (t,x) \in \R \times [0,\ell_n].
$$
Then, thanks to 
\eqref{bepsuno}, \eqref{aepsuno} we have 
$$
\langle{\spaceimm_n}_t,
{\spaceimm_n}_x\rangle=0,  \qquad
\vert {\spaceimm_n}_t\vert^2 
+ \vert {\spaceimm_n}_x\vert^2=1
$$
and the wave system \eqref{wavelinearsystem}. 
Now we claim 
 that
 there exists $\alpha >1$ such that for any $s \in \R$  
\begin{equation*}
\displaystyle
\lim_{
n \to +\infty
}
a_n(s) = \alpha a\left(\frac{s}{\alpha}\right), \qquad
\qquad  \displaystyle
\lim_{n \to +\infty
}
b_n(s) =  
a\left(\frac{s}{\alpha}\right).
\end{equation*}
To prove the claim, let $\phi(p) :=
\displaystyle \sqrt{\frac{5}{4}+p}$ for any $p \geq - \frac{5}{4}$, 
and observe that
\begin{equation}\label{conv_int_osc}
\lim_{n \to +\infty
} s_n(x) =
\frac{x}{2\pi} \int_0^{2\pi} \phi(\cos \varmuta
)~ d\varmuta. 
\end{equation}
Indeed, $s_n(x) = 
 \int_0^x \phi(\cos(n\varmuta - \varmuta))~ d\varmuta$, and
the change of variable $y=n\varmuta
-\varmuta$ gives
$$
s_n(x) = \frac{x}{x(n-1)} 
\int_0^{x(n-1)} \phi(\cos y)~ dy.
$$
Hence $s_n(x)$ equals $x$ times 
the mean value of the $2\pi$-periodic function
$\phi(\cos y )$ in the interval $[0,x(n-1)]$. 
We now claim that such a mean value 
converges to the mean value of 
$\phi(\cos y)$ 
on $[0,2\pi]$. Indeed, denoting by $[r]$
the integer part of $r \in \R$, we have, for $x>0$, 
\begin{equation}\label{secondremarkjens}
\frac{s_n(x)}{x} = \frac{1}{x(n-1)}
\int_0^{
2\pi [\frac{x(n-1)}{2\pi}]
}
\phi(\cos \sigma)~d\sigma  + 
\frac{1}{x(n-1)}
\int_{
2\pi [\frac{x(n-1)}{2\pi}]
}^{x(n-1)}
\phi(\cos \sigma)~d\sigma,
\end{equation}
and the last addendum on the right hand side
converges to zero as $n \to +\infty$. The claim 
then follows, since the denominator of the first addendum 
on the right hand side of \eqref{secondremarkjens} reads 
as $x(n-1) = 
2\pi [\frac{x(n-1)}{2\pi}] 
+  o_n$, where 
$o_n := x(n-1) -
2\pi [\frac{x(n-1)}{2\pi}]$ converges to zero
as $n \to +\infty$.

{}From the claim we conclude that 
formula \eqref{conv_int_osc} holds.
Define now
\begin{equation}\label{alfa}
\alpha := \frac{1}{2\pi} \lim_{n \to +\infty
} s_n(2\pi) = 
\frac{1}{2\pi} \int_0^{2\pi}
\sqrt{\frac{5}{4}+\cos \sigma}~d\sigma >1,
\end{equation}
so that $\lim_{n \to +\infty} s_n(x) = \alpha x$,  hence 
$\lim_{\eps \to 0} x_n(s) = s/\alpha$, and the claim
follows.

Then 
$$
\displaystyle
\lim_{n \to +\infty}\spaceimm_n(t,x) =
\frac{1}{2} \left[\alpha a\left(\frac{x+t}{\alpha}\right) 
+ a\left(\frac{x-t}{\alpha}\right)\right]=: \spaceimm(t,x)
$$
uniformly for $(t,x)$ in the compact subsets of $\R \times \R$.

The limit curve $\spaceimm$ is such that $\spacetimeimm(t,x) :=
(t,\spaceimm(t,x))$ is not a critical point of $\area$; it is interesting to
observe, as remarked in \cite{Ne:90},
 that the additional oscillations ``desingularize'' the limit,
in the sense that the image of the map $\spaceimm$ has not
anymore any singular point.
\end{Example}

The last example is similar to Example \ref{neuoscillating}, but 
in $n=3$ dimensions; here the situation is 
simpler, since the analog of the arc-length reparametrization
in \eqref{bepsarclength} is automatically satisfied. 

\begin{Example}\label{exa:giandomenico}\rm
Assume $n=3$.
Consider cylindrical coordinates in $\R^3$
and set, for $\paramgiando \in \R$,
$$
e_r:=(\cos\paramgiando
,\sin\paramgiando
,0),
\qquad e_\paramgiando
:=(-\sin\paramgiando
,\cos\paramgiando
,0)\, ,\qquad e_z:=(0,0,1).
$$
Let $\alpha, \beta \in (-1,1)$ be such that 
$\alpha^2+\beta^2=1$, $n \in \N$, and
define the $2\pi$-periodic maps $a, b_n: \R\to \R^3$ as
\begin{equation*}
\begin{aligned}
a(\paramgiando
) := &
e_\paramgiando
, 
\\
 b_n(\paramgiando
):= &\alpha  e_\paramgiando
 +\beta 
\left( e_\paramgiando
\sin(n\paramgiando
)\frac{n}{n^2-1}
 - e_r\cos(n\paramgiando
)\frac{1}{n^2-1}
 + e_z\cos(n\paramgiando
)\frac{1}{n}\right).
\end{aligned}
\end{equation*}
A direct computation gives
$$
\begin{aligned}
b'_n(\paramgiando
) &=-\alpha
 e_r -\beta  e_r \sin(n\paramgiando
)\frac{n}{n^2-1}
-\beta e_\paramgiando
 \cos(n\paramgiando
)\frac{1}{n^2-1} \\
&{\quad\ }+\beta e_\paramgiando
\cos(n\paramgiando
)\frac{n^2}{n^2-1}+\beta
 e_r\sin(n\paramgiando
)\frac{n}{n^2-1}-\beta
e_z\sin(n\paramgiando
)\\
&=-\alpha  e_r +\beta 
 e_\paramgiando
\cos(n\paramgiando
)-\beta  e_z\sin(n\paramgiando
).
\end{aligned}
$$
so that 
$$
|b'_n(\paramgiando
)|^2=\alpha^2+\beta^2=1, \qquad \vert a'(\paramgiando
)\vert^2=1,
\qquad \paramgiando
\in \R.
$$
Moreover 
$$
\lim_{n\to +\infty}
b_n(\paramgiando
) = \alpha  e_\paramgiando
=\alpha  a(\paramgiando
) =: b(\paramgiando
)
$$
uniformly in $\R$.

Define
$$
\spaceimm_n(t,\paramgiandox):=\frac{1}{2} \left[a(\paramgiandox
+t)+b_n(\paramgiandox
-t)\right], \qquad (t,\paramgiandox) \in \R \times \R.
$$
Then $\spaceimm_n$ satisfy $\langle {\spaceimm_n}_t, 
{\spaceimm_n}_\paramgiandox\rangle =0$, $\vert 
{\spaceimm_n}_t\vert^2 + \vert {\spaceimm_n}_\paramgiandox 
\vert^2 =1$, and \eqref{wavelinearsystem}. Moreover
$$
\lim_{n \to +\infty}
\spaceimm_n(t,\paramgiandox) =
\frac{1}{2} \left[
a(\paramgiandox
+t)+\alpha
 a(\paramgiandox
-t)\right]
=:
\spaceimm(t,\paramgiandox)
$$
uniformly in on the compact subsets of $\R \times \R$. 
Also in this example $\spaceimm(t,\paramgiandox)$ cannot vanish, since
$$
|\spaceimm(t,\paramgiandox)|^2=\frac{1}{4}
\left[
1+\alpha^2+2\alpha
\cos (2t)\right]=
\frac{1}{4}
\left[(1+\alpha)^2\cos^2t+(1-\alpha)^2\sin^2t\right]
\geq
\frac{(1-\alpha)^2}{4}>0.
$$
Observe that letting $a(\paramgiando
)=(-\sin(\paramgiando
+2\phi),
\cos(\paramgiando
+2\phi),0)$ for $\phi\in (0,\pi)$, we have for the resulting $\spaceimm$
$$
|\spaceimm(t,\paramgiandox)|^2=
\frac{1}{4} \left[
(1+\alpha)^2\cos^2(t+\phi)+(1-\alpha)^2\sin^2(t+\phi)\right],
$$
and again $|\spaceimm(t,\paramgiandox)|\ge (1-\alpha)/2$.
\end{Example}

It would be interesting to understand whether there are connections
between the examples considered in this section 
and the results of \cite{FrMu:92}.

\section{Evolution of $\C^2$ uniformly convex curves with 
$\spaceimm_t(0,\cdot)=0$}\label{sec:convex}
Let $\overline t >0$ and let 
$\spaceimm \in C^2([0,\overline t) \times [0,
\energy
];\Rn)$ be a solution of 
 \eqref{tang_vect_mutually_orthogonal}, \eqref{secondovincolo}
and \eqref{wavelinearsystem}.
In particular, there exist $\energy$-periodic maps 
$a,b\in C^2(\R;\Rn)$ such that $\spaceimm(t,x) 
= \frac{1}{2}\left[a(x+t)+b(x-t)\right]$ for any $(t,x) \in 
[0,\overline t) \times [0,E]$. 
Therefore, recalling the discussion in Remark \ref{rem:mettitogli}, 
$\spaceimm$ can be extended to a global solution 
$\weaksol \in \C^2(\R \times [0,\energy];\Rn)$. 
Adopting this definition of global solution, we show in this section 
that initial convex curves may shrink to a point,
and then continue the motion in a periodic way.

\begin{Definition}\label{def:extinction}
Let $\overline t>0$ and $p \in \Rn$. 
We say that $\overline t$ is a collapsing time, and that 
$\weaksol$ has a collapsing singularity 
at $\overline t$ with $p$ as collapsing point, if
 $\weaksol(\overline t,x) = p$
for any $x \in [0,\energy]$. 
\end{Definition}

At the collapsing time we have 
\begin{equation}\label{cond:extinction}
0=\weaksol_x(\overline t,x) = \frac{1}{2}
\left[
a'(x+\overline t) + b'(x-\overline t)\right], \qquad x \in [0,\energy].
\end{equation}
Let us now assume $n=2$, $\spaceimm_t(0,\cdot)=0$, so that 
we can choose $a=b \in \C^2(\R; \R^2)$.  We also assume that
 $a$ 
parametrizes, on $[0,\energy]$, a closed uniformly convex curve
of class $\C^2$.
Since the initial curve is uniformly convex, 
for any $x\in [0,\energy]$ 
there exists a unique ${\mathrm t}(x) \in (0,\energy/2)$ such that 
\begin{equation}\label{def:trm}
\spaceimm_x({\mathrm t}(x),x) = \frac{1}{2} \left[
a'(x+{\mathrm t}(x)) + a'(x-{\mathrm t}(x))\right] =0,
\end{equation}
and the function ${\mathrm t}$ belongs to $\C^1([0,\energy];(0,\energy/2))$.
Moreover, if we set
\[
{\mathrm t}_{\rm min} := \min_{x\in [0,\energy]}{\mathrm t}(x)
\qquad 
{\mathrm t}_{\rm max} := \max_{x\in [0,\energy]}{\mathrm t}(x),
\]
we have 
 that $\spaceimm(t,\cdot)$ is a regular parametrization for all $t\in 
[0,{\mathrm t}_{\rm min})\cup ({\mathrm t}_{\rm max},\energy/2]$.
We can think of ${\mathrm t}_{\min}$ (resp. ${\mathrm t}_{\max}$) as the 
first (resp. last) singularity time in the periodicity
interval $[0,\energy]$, where by singularity here we mean
that the regularity condition of Definition \ref{def:regular} fails.
\begin{Proposition}\label{prop:whereby}
Let $\spaceimm \in \C^2([0,{\mathrm t}_{\rm min})\times [0,\energy]; \R^2)$
 be a solution
of \eqref{eq:EL} given by \eqref{eq:representgamma}.
Assume that  $\spaceimm(0,\cdot) \in \C^2([0,\energy])$
is regular and embedded, that $\spaceimm(0,[0,\energy])$  
encloses a compact centrally symmetric
uniformly convex body $K(0)$, and that $\spaceimm_t(0,\cdot)=0$.
Then $\weaksol$ has a collapsing singularity at time 
$t_{\min}=\energy/4$ with the origin as collapsing point.
\end{Proposition}
\begin{proof}
The assertion follows by observing that $K(0)$ is centrally symmetric,
and  the function ${\mathrm t}$ defined in \eqref{def:trm}
is constant and equals
$\energy/4 = t_{\min}$.
\end{proof}

\begin{Remark}\label{rem:persist}\rm
Generically, one can assume that
\begin{itemize}
\item[-]
the last equality in
 \eqref{cond:extinction} does not hold;
\item[-]
the set $\{x\in [0,\energy]:\,{\mathrm t}(x)=t\}$
is finite for all $t\in [{\mathrm t}_{\rm min},{\mathrm t}_{\rm max}]$, 
and consists of a single point 
$x_{\rm min}$ (resp. $x_{\rm max}$) for $t={\mathrm t}_{\rm min}$ (resp. 
$t={\mathrm t}_{\rm max}$).
\end{itemize}
{}From the condition ${\mathrm t}'(x_{\rm min})={\mathrm t}'(x_{\rm max})=0$ we get
\[
a''(x_{\rm min}+{\mathrm t}_{\rm min})=-a''(x_{\rm min}-{\mathrm t}_{\rm min}) 
\qquad {\rm and}\qquad 
a''(x_{\rm max}+{\mathrm t}_{\rm max})=-a''(x_{\rm max}-{\mathrm t}_{\rm max}),
\]
which implies that the images
$\spaceimm({\mathrm t}_{\rm min},[0,E])$ and $\spaceimm(
{\mathrm t}_{\rm max},[0,E])$
are of class $\C^1$.
In this generic setting, 
the formation of singularities has been discussed in \cite{EH} 
(see also \cite{ViSh:94}, \cite{An:03}),
where it is shown that ${\mathrm t}_{\rm min}$ is the first singular time, the
singularity 
has the asymptotic behavior $y\sim x^\frac{4}{3}$ in graph coordinates, and two
cusps $y\sim x^\frac{2}{3}$ appear 
from the point $x_{\rm min}$ at time ${\mathrm t}_{\rm min}$, 
persist for some positive time, and eventually disappear. 
\end{Remark}

We now show that the convexity of the curve is preserved before the 
onset of singularities, that is on the time interval 
$[0,{\mathrm t}_{\rm min})$.
\begin{Proposition}\label{proconvex}
Let $\spaceimm \in \C^2([0,{\mathrm t}_{\rm min})\times [0,\energy]; \R^2)$ be a
solution
of \eqref{eq:EL} given by \eqref{eq:representgamma}.
Assume that  $\spaceimm(0,\cdot) \in \C^2([0,\energy])$
is embedded and counter-clockwise regularly parametrized,
that $\spaceimm(0,[0,\energy])$  
encloses a compact
uniformly convex body $K(0)$, and that $\spaceimm_t(0,\cdot)=0$.
Then $\spaceimm(t,\cdot)$ is the regular parametrization of a 
closed uniformly convex embedded curve of class $\C^2([0,\energy])$
for all $t\in [0,{\mathrm t}_{\rm min})$. Moreover, letting $K(t)$ the 
compact convex set enclosed by 
$\spaceimm(t,\cdot)$, we have
\begin{equation}\label{eq:inclusionprinciple}
t_1, t_2 \in [0,{\mathrm t}_{\rm min}), ~ t_1 \leq t_2 \quad \Rightarrow
\quad K(t_1) \subseteq K(t_2).
\end{equation}
%
\end{Proposition}
\begin{proof}
For any $t \in [0,t_{\min})$ 
let 
\[
\nu(t,x) = R\,\frac{\spaceimm_x(t,x)}{|\spaceimm_x(t,x)|}, \qquad 
(t,x)\in [0,{\mathrm t}_{\rm min})\times [0,\energy],
\] 
where $R:\R^2\setminus \{0\}\to \R^2\setminus \{0\}$ is the 
counter-clockwise rotation of $\pi/2$.
To prove that $\spaceimm(t,\cdot)$ is a uniformly convex curve, 
it is 
enough to show that 
$$
\langle \spaceimm_{xx}(t,\cdot),\nu(t,\cdot)\rangle > 0, \qquad
t \in [0,{\mathrm t}_{\rm min})\times [0,\energy].
$$
{}From $\spaceimm_x(t,\cdot)\ne 0$ for $t \in [0,{\mathrm t}_{\min})$ it
follows that 
\begin{equation}\label{sumnezero}
a'(x+t)+a'(x-t) \ne 0, \qquad (t,x)\in [0,{\mathrm t}_{\rm min})\times [0,
\energy].
\end{equation}
Hence
\begin{eqnarray*}
\langle \spaceimm_{xx}(t,x),\nu(t,x)\rangle &=& 
\frac{1}{2} 
\langle a''(x+t) + a''(x-t), \frac{Ra'(x+t)+Ra'(x-t)}{\vert
a'(x+t)+a'(x-t)\vert}\rangle.
\end{eqnarray*}
Observe now that $\vert a'\vert=1$ implies that 
$a''(x\pm t)\perp a'(x\pm t)$, so that $a''(x\pm t)$ 
and $R a'(x\pm t)$ are parallel. Then \eqref{sumnezero} 
and  the Schwarz
inequality imply that 
\begin{eqnarray*}
\langle a''(x+t) , Ra'(x+t)+ Ra'(x-t) \rangle &>& 0
\\
\langle a''(x-t), Ra'(x+t)+ Ra'(x-t) \rangle &>& 0,
\end{eqnarray*}
which gives 
\[
\langle \spaceimm_{xx}(t,x),\nu(t,x)\rangle > 0.
\]
It remains to prove \eqref{eq:inclusionprinciple}. 
Equation
\eqref{eqgeom} and the uniform convexity $\langle \kappa,\nu\rangle>0$ 
imply 
$$
\langle {\rm a},\nu\rangle = 
(1-\vert v\vert^2) \langle \kappa,\nu\rangle
>0.
$$
Recalling that $\spaceimm_t(0,\cdot)=0$ and that 
$\langle \mathrm a, \nu \rangle = \partial_t
\langle v, \nu\rangle>0$, we 
then get $\langle {v},\nu\rangle>0$ for any 
$t \in (0,t_{\min})$,
and \eqref{eq:inclusionprinciple} follows.
\end{proof} 

A result analogous to 
Proposition \ref{proconvex} has been obtained in \cite{KLW} 
for the equation 
${\mathrm a}  = \kappa$.
Differently from our case, 
for their equation the authors of 
\cite{KLW} show that all convex curves shrink to a point in finite time.
\begin{Remark}\label{rem:extinctionform}\rm
Assume (as in Proposition \ref{prop:whereby})
that the initial uniformly convex set is of class $\C^{2}$,
and that $\spaceimm$ has a collapsing singularity
at the time $\overline t = \energy/4$, with $p\in \R^2$ as 
collapsing point. 
{}From the representation
formula 
\eqref{eq:representgamma} with $a=b$, and from
Taylor's formula,
we get
\begin{eqnarray*}
\spaceimm(t,x) &=& 
\frac{1}{2}\left[
a(x+\overline t)+a(x-\overline t)\right] + \frac{1}{2}
\left[a'(x-\overline
t)-a'(x+\overline t)\right]\, (\overline t-t)
+ O(|\overline t -t|^2)
\\
&=& p + a'(x-\overline t)\, (\overline t -t) + O(|\overline t -t|^2),
\end{eqnarray*}
where in the last equality we use $a'(x + \overline t) 
+ a'(x -\overline t)=0$ (see \eqref{def:trm}).
It follows that
\begin{equation}\label{eqcirc}
\vert \spaceimm(t,x)-p \vert = \vert
\overline t-t\vert + O(\vert \overline t -t\vert^2).
\end{equation}
In particular, 
the asymptotic shape near the collapse is circular,
and the blow-up shape of the image of the 
corresponding map $\spacetimeimm$ (see \eqref{gammagamma})
 at $(\overline t,p)$ 
is half a light cone.
\end{Remark}

The conclusion on the asymptotic shape of $\spaceimm$
in 
Remark \ref{rem:extinctionform} seems not to be true if we drop the $\C^{1,1}$
regularity assumption on the initial convex set, as shown in the following
example. 

\begin{Example}\label{ex:square}\rm
Assume $n=2$, let $L>0$ and let $a = b : \R \to \R^2$
be $4L$-periodic, and such that $a: [0,4\rightextremum]
\to \R^2$
be the counterclock-wise arc-length parametrization of the boundary of 
the square $Q_0 = [-L/2,L/2]^2$
(sending for instance $\{0\}$ into the point 
$x^1 = -L/2$, $x^2=-L/2$).
Obviously $a \in \C^1([0,4\rightextremum] \setminus \{0,
\rightextremum,2 \rightextremum,3 \rightextremum\}; \R^2)$, and 
$a$ is Lipschitz continuous in $[0,4\rightextremum]$.

Then, letting $\spaceimm(t,x):= \frac{1}{2}\left[
a(x+t) + a(x-t)
\right]$ for any $(t,x) \in \R \times \R$,
we have that $\spaceimm(t,\cdot)$ is a Lipschitz parametrization of 
$\partial Q(t)$,
where $Q(t)$ is defined as
\[
Q(t) := Q_0\cap \left\{ (x^1,x^2)\in\R^2:\,|x^1|+|x^2|\le L-t \right\},
\qquad t\in [0,\rightextremum].
\]
For times larger than $L$ the solution is continued periodically, hence 
$\spaceimm$ is Lipschitz in $\R \times [0,E]$, and therefore
it is almost everywhere differentiable. 
\begin{figure}
\begin{center}
\includegraphics[height=4.5cm]{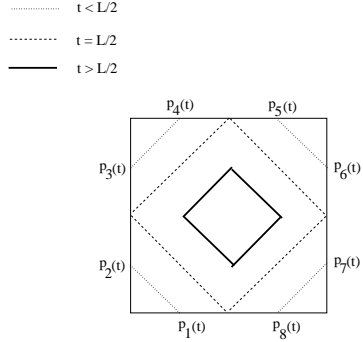}
\caption{\small A (weak) evolution of the square with
zero initial velocity.
} 
\label{fig:square}
\end{center}
\end{figure}
Observe that 
\begin{itemize}
\item[(i)] the map $\spacetimeimm(t,x) :=(t, \spaceimm(t,x))$
is Lipschitz, and at those points of 
$\spacetimeimm(\R\times [0,4\rightextremum])$ where there exists
the tangent plane such a plane is time-like. 
\item[(ii)] 
For $t\in [0,L/2)$ the set
$Q(t)$ is a shrinking octagon, with   
vertices $p_1(t), \ldots, p_8(t)$ (see Fig. \ref{fig:square}). 
For this interval of times the conservation 
law \eqref{cons:image} is satisfied, 
since 
$$
\int_{\spaceimm(t,[0,4\rightextremum])} \frac{1}{\sqrt{1-\vert
\spaceimm_t^\perp\vert^2}}~d\mathcal H^1
= 4 \left[ 
\vert p_8(t) - p_1(t)\vert + \sqrt{2}
\vert p_8(t) - p_7(t) \vert
\right] = 4L.
$$
Moreover, for $t \in [0,L/2)$ the map $\spaceimm$ is strictly admissible,
in the sense that  
$\vert \spaceimm^\perp\vert^2<1$ almost everywhere.
\item[(iii)]  For $t\in [L/2,L)$ the set
$Q(t)$ is a shrinking rotated square
of side $\sqrt 2 (L-t)$
(depicted in bold in Fig. \ref{fig:square}).
It shrinks to
the point $(0,0)$ at $t=L$ (collapsing
singularity).
Its normal
velocity is constantly equal to $\frac{1}{\sqrt{2}}$.
Therefore \eqref{cons:image}
{\it cannot} be satisfied, since the time derivative of the length
of $\spaceimm(t,\cdot)$ is nonzero. However, 
the function 
$$
t \in [L/2,L)\to \int_{\spaceimm(t,[0,4\rightextremum])} 
~ \frac{1}{\sqrt{1 - 
\vert \spaceimm_t^\perp(t,\cdot)\vert^2
}} ~d \mathcal H^1
$$
is nonincreasing.
\item[(iv)] Given $t \in (L/2,L)$,  we have $\spaceimm_x(t,x)=0$ 
when $x$ belongs to the union $I(t)$ 
of four  
intervals of length $2t-L$, and  centered at the centers of the four
sides of $\partial Q_0$. Indeed, $\spaceimm_x(t,x)=0$ when 
$a'(x+t)=-a'(x-t)$, hence, for instance
assuming $x$ to be the center of $[-L/2,L/2] \times \{-L/2\}$,
when $x+t$ and $x-t$ belong to opposite vertical sides
of $\partial Q_0$.
Therefore,
for $t \in (L/2,L)$ and $x \in I(t)$, we have that 
$\spaceimm(t,\cdot)$ is not regular, 
$$
\lagdens(\spaceimm_t(t,x),\spaceimm_x(t,x))=0,
\qquad
\vert \spaceimm_t(t,x)\vert^2 =1,
$$
and \eqref{strictlytimelike} is not satisfied.
\end{itemize}
Note that the blow-up of $\spacetimeimm$
at $(L,0)$ is not half a light cone as in Remark \ref{rem:extinctionform},
but is the half-cone 
$\{(t,x_1,x_2) : \vert t-\overline t\vert + \vert
x_1\vert + \vert x_2\vert =1\}$
with square section, 
inscribed in half the light-cone.

\end{Example}

\noindent {\bf Acknowledgements.}
We would like to thank the Swedish Research Council,
and KTH, for support.


\begin{thebibliography}{99}

\bibitem{Am:98}
L. Ambrosio,
\newblock Geometric evolution problems, distance function and 
viscosity solutions.
\newblock In: {\em Calculus of Variations and Partial Differential
Equations. Topics on Geometrical Evolution Problems and
Degree Theory}, Springer-Verlag, 1999.

\bibitem{AmFuPa:00}
L. Ambrosio, N. Fusco, D. Pallara,
{\em Functions of Bounded Variation
                  and Free Discontinuity Problems}.
Clarendon Press (Oxford), 2000.

\bibitem{An:03}
M.A. Anderson,
\newblock {\em 
The Mathematical Theory of Cosmic Strings. Cosmic Strings
 in the Wire Approximation.}
\newblock Institute of Physics Publishing,
Briston and Philadelphia, 2003.


\bibitem{BeNoOr:09}
G. Bellettini, M. Novaga, G. Orlandi.
\newblock Time-like minimal submanifolds as singular limits of nonlinear
wave equations.
\newblock {\em Physica D}, to appear.

\bibitem{Br:02}
S. Brendle. 
\newblock Hypersurfaces in Minkowski space with vanishing mean curvature. 
\newblock {\em Comm. Pure Appl. Math.}, 55(10):1249?-1279, 2002.

\bibitem{Br:05}
Y. Brenier
\newblock Non relativistic strings may be approximated by 
relativistic strings.
\newblock {\em Methods Appl. Anal}, 2005.

\bibitem{BoIn:34}
M. Born, L. Infeld.
\newblock Foundations of a new field theory.
\newblock {\em Proc. Roy. Soc. A}, 144:425--451, 1934.

\bibitem{Ca:95}
B. Carter.
\newblock Dynamics of cosmic strings and other brane models.
\newblock {In {\em Formation and interactions of topological defects}, NATO Adv.
Sci. Inst. Ser. B Phys.},
349:303--348, (1995).

\bibitem{EH}
J.~Eggers, J.~Hoppe.
\newblock Singularity formation for time-like extremal
hypersurfaces.
\newblock {\em Physics Letters B}, 680 (2009), 274--278.

\bibitem{FrMu:92}
G.A.~Francfort, F.~Murat.
\newblock Oscillations and energy densities in the wave equation.
\newblock {\em Commun. Partial Differ. Equations} 17:1785-1865, 1992.

\bibitem{GaHa:86}
M. Gage, R.S. Hamilton.
\newblock The heat equation shrinking convex plane curves.
\newblock {\em J. Differential Geom.} {\bf 23} (1986),
69--96.

\bibitem{GiIs:04}
G.W. Gibbons, A. Ishibashi.
\newblock Topology and signature in braneworlds. 
\newblock {\em Class. Quantum Grav.}, 21:2919-2935, 2004.

\bibitem{Gr:87}
M.A. Grayson.
\newblock The heat equation shrinks embedded plane curves to round points.
\newblock {\em J. Differential Geom.}, {\bf 26} (1987),
285--314

\bibitem{Ho:1}
J. Hoppe.
\newblock Membranes and matrix models.
\newblock hep-th/0206192, IHES/P/02/47.

\bibitem{Ho:2}
J. Hoppe.
\newblock Conservation laws and formation of singularities 
in relativistic theories of extended objects.
\newblock hep-th/9503069.

\bibitem{Li:04}
H.~Lindblad.
\newblock A remark on global existence for small initial data
of the minimal surface equation in Minkowskian space time.
\newblock {\em Proc. Am. Math. Soc.}, 132(4):1095--1102, 2004.

\bibitem{KLW}
D.X. Kong, L. Kefeng, Z.G. Wang.
\newblock Hyperbolic mean curvature flow: evolution of plane curves.
\newblock {\em Acta Math. Sci. Ser. B Engl. Ed.}, 29(3):493--514, 2009.

\bibitem{milbredt}
O. Milbredt.
\newblock The Cauchy problem for membranes.
\newblock {\em         arXiv:0807.3465v1}, 2008.

\bibitem{Ne:90}
J.C.~Neu.
\newblock Kinks and the minimal surface equation in Minkowski space.
\newblock {\em Physica D}, 43(2-3):421--434, 1990.

\bibitem{RN:00}
H. Rotstein, A. Nepomnyashchy.
\newblock Dynamics of kinks in two-dimensional hyperbolic models.
\newblock {\em Physica D}, 136(3-4):245--265, 2000.

\bibitem{ViSh:94}
A. Vilenkin, E.P. S. Shellard.
\newblock {\em Cosmic Strings and Other Topological Defects.}
\newblock Cambridge University Press, 1994.

\bibitem{Zw:04}
B. Zwiebach.
\newblock {\em A First Course in String Theory.}
\newblock Cambridge University Press, second edition, 2009.


\end{thebibliography}
\end{document}